\documentclass[journal]{IEEEtran}
\usepackage{graphicx,url}
\usepackage{float}
\usepackage[caption = false]{subfig}
\usepackage[justification=centering]{caption}
\usepackage{tikz}
\usepackage{textcomp}
\usepackage{hyperref}
\usepackage{lipsum}

\newcommand\copyrighttext{%
  \footnotesize \textcopyright 2016 IEEE. Personal use of this material is permitted.
  Permission from IEEE must be obtained for all other uses, in any current or future
  media, including reprinting/republishing this material for advertising or promotional
  purposes, creating new collective works, for resale or redistribution to servers or
  lists, or reuse of any copyrighted component of this work in other works.
  DOI: 10.1109/TASLP.2016.2592698,}
\newcommand\copyrightnotice{%
\begin{tikzpicture}[remember picture,overlay]
\node[anchor=south,yshift=8pt] at (current page.south) {\fbox{\parbox{\dimexpr\textwidth-\fboxsep-\fboxrule\relax}{\copyrighttext}}};
\end{tikzpicture}%
}

\ifCLASSINFOpdf
\else
\fi

\makeatletter

\newcommand{\Rmnum}[1]{\expandafter\@slowromancap\romannumeral #1@}
\makeatother

\begin{document}
%
\title{Automatic Environmental Sound Recognition: Performance versus Computational Cost}
%
%
%

\author{Siddarth~Sigtia,~\IEEEmembership{Student Member,~IEEE,}
        Adam~M.~Stark,\\
        Sacha~Krstulovi{\'c} 
        and~Mark~D.~Plumbley,~\IEEEmembership{Fellow,~IEEE}
\thanks{S. Sigtia (s.s.sigtia@qmul.ac.uk) and A. Stark (adamstark.uk@gmail.com) were with Queen Mary University of London during
the development of this research work. 
S.~Krstulovi{\'c}  (sacha.krstulovic@audioanalytic.com) is with Audio
Analytic Ltd. 
Prof.~M.~D.~Plumbley (m.plumbley@surrey.ac.edu) is with the University of
Surrey, Guildford.}
\thanks{Manuscript received MONTH DAY, 2015; revised MONTH DAY, 2015.}}

\maketitle
\copyrightnotice
\begin{abstract}
In the context of the Internet of Things (IoT), sound sensing applications are
required to run on embedded platforms where notions of product pricing and form
factor impose hard constraints on the available computing power. Whereas
Automatic Environmental Sound Recognition (AESR) algorithms are most often developed
with limited consideration for computational cost, this article seeks which
AESR algorithm can make the most of a limited amount of computing
power by comparing the sound classification performance {\em as a function of
its computational cost}. Results suggest that Deep Neural Networks yield
the best ratio of sound classification accuracy across a range of computational
costs, while Gaussian Mixture Models offer a reasonable accuracy at a
consistently small cost, and Support Vector Machines stand between both in
terms of compromise between accuracy and computational cost.
\end{abstract}


%
\IEEEpeerreviewmaketitle

\section{Introduction \label{sec:intro}}
\IEEEPARstart{A}{utomatic} speech recognition, music classification, audio
indexing, and to a less visible extent biometric voice authentication, have now
achieved some degree of commercial success in mass markets. A vast proportion
of these applications are supported by PC platforms, cloud computing and
powerful smartphones. Nevertheless, a new market is quickly emerging in the
domain of the {\em Internet of Things} (IoT) \cite{gubbi2013internet}. In this market, Automatic
Environmental Sound Recognition (AESR) \cite{chachada2014environmental} 
has a significant potential to create value, e.g., for security 
or home safety applications \cite{istrate2006information,vacher2010challenges,sitte2007non}.

However, in this context, industrial reality imposes hard constraints on the
available computing power. Indeed, IoT devices can be broadly divided into
two classes: (a) devices which are perceived as ``doing only one thing'', thus
requiring the use of low cost processors to hit a price point that users are
willing to pay for what the device does; (b) embedded devices where sound
recognition comes as a bolt-on to add value to an existing product, for
example adding audio analytic capabilities to a consumer grade camera or to a
set-top box, thus requiring the algorithm to fit into the device's existing
design and price points. These two cases rule out the use of higher end processors.
Generally speaking, the following features jointly define the financial cost of
a processor and the level of constraint imposed on embedded computing: a) the {\em clock
speed} is related to energy consumption; b) the {\em instruction set} is related to
chip size and manufacturing costs, but in some cases includes special
instruction sets to parallelise more operations into a single clock cycle;
c) the {\em architecture} defines, e.g., the number of registers, number of cores,
presence/absence of a Floating Point Unit (FPU), a Graphical Processing Unit
(GPU) and/or a Digital Signal Processing (DSP) unit. These features affect applications
in the obvious way of defining an upper limit on the number and the type of
operations which can be executed in a given amount of time, thus ruling
the possibility to achieve real time audio processing at the proportion of
processor load allocated to the AESR application. Finally, {\em onboard memory size} is
an important factor related to processor cost, as it affects
both the computational performance, where repetitive operations can be cached
to trade speed against memory, and the scalability of an algorithm, by imposing
upper limits on the number of model parameters that can be stored and manipulated.

If trying to work within these limitations, and given that most IoT embedded
devices allow Internet connectivity, it could be argued that cloud computing can
solve the computational power constraints by abstracting the computing platform
and making it virtually as powerful as needed. However, a number of additional
design considerations may rule out the use of cloud computing for
AESR applications: the {\em latency} introduced by cloud communications can be
a problem for, e.g., time critical security applications \cite{bonomi2014fog}; regarding {\em
Quality of Service (QoS)}, network interruptions introduce an extra point of
failure into the system; regarding {\em privacy} and {\em bandwidth
consumption}, respectively, sending alerts rather than streaming audio or
acoustic features out of the sound recognition device both rules out any
possibility of eavesdropping~\cite{medaglia2010overview} and requires less
bandwidth.

Thus, the reality of embedded industrial applications is that at the price
points acceptable in the marketplace, IoT devices will most likely be devoid of
an FPU, will operate in the hundreds of Megahertz clock speed range, won't offer
onboard DSP or specialised instruction sets, and may prefer to run AESR on board
rather than in the cloud. At the same time, most AESR performance evaluations
are still produced with limited interest for the computational cost involved:
many experimental results are obtained with floating point arithmetics on
powerful computing platforms, and under the assumption that sufficient
computing power will sooner or later become available for the algorithm to be
practicable in the context of commercial applications.

This article aims at crossing this divide, by comparing the performance
of three AESR algorithms as a function of their computational cost. In other
terms, it seeks which AESR algorithm is able to make the most of a
limited amount of computational power. The rest of the paper is organised as follows:
Section~\ref{sub:prior} reviews prior art in AESR. Section~\ref{sec:cost}
defines the measure of computational cost used in this study, and
applies it to a range of popular machine learning algorithms.
Section~\ref{sec:perf} defines the other axis of this evaluation, which is the
metric used to measure classification performance. The data sets and methodology
are detailed in Section~\ref{sec:methodology}, while results are discussed in
Section~\ref{sec:results}.

\section{Background \label{sub:prior}}
Audio or sound recognition research has primarily been focused on 
speech and music signals. However in the past decade, the problem of classifying 
and identifying environmental sounds has received more attention \cite{chachada2014environmental}. 
AESR has applications in content-based search \cite{virtanen2007probabilistic}, robotics 
\cite{yamakawa2011environmental}, security \cite{istrate2006information} and a host of other tasks. 
Recognition of environmental sounds is a part of the much broader field of computational 
auditory scene analysis (CASA) \cite{wang2006computational}. We are interested in the problem 
of identifying the presence of non-speech, non-music sounds or \emph{environmental sounds}
in a given audio recording. Environmental sound recognition involves the study of two closely related 
problems. Acoustic Scene Classification (ASC) involves selecting an appropriate set of labels 
that encompasses all the sound events which occur in a recording \cite{stowell2015detection}, e.g., ``crowded train station''. With regards to machine
learning, ASC can be seen as a sequence classification task where a set 
of labels is assigned to a variable length input sequence. A related problem is the problem 
of Acoustic Event Detection (AED). In addition to identifying the correct label, AED aims to 
segment the audio recording into regions where the label occurs \cite{stowell2015detection}, e.g., ``stationmaster blowing whistle'' or ``doors closing''. Again with regards to machine 
learning, AED involves identifying the correct labels and the correct \emph{alignment}
of labels with the recording. In this study, we compare Gaussian mixture models (GMMs), 
support vector machines (SVMs) and deep neural networks (DNNs) in terms of their 
performance on an AED task as a function of their computational cost.

GMMs are generative probabilistic models
 and have been extensively applied to many problems in speech
recognition~\cite{rabiner1993fundamentals}. In a number of cases, GMMs have
therefore been considered a sensible starting point for research on the
classification of environmental sounds. For example, they have been used to
detect gunshot sounds~\cite{clavel2005events} and to discriminate between
gunshots and screams~\cite{valenzise2007scream}. Port\^{e}lo et
al.~\cite{portelo2009non} have used Gaussian Mixtures as part of a hidden
Markov model (HMM) approach to detecting sirens in audio. Elsewhere,
Radhakrishnan et al.~\cite{radhakrishnan2005audio} have used GMMs when
attempting to develop a model for recognising new `outlier' sounds against a
GMM trained to model background sounds. Principi et al.~\cite{Principi2015}
proposed another novelty detection system, with an embedded implementation on a
Beagle Board~\cite{Bonfigli2014}. Atrey et al.~\cite{atrey2006audio} have used
GMMs to classify audio frames into foreground or background sounds, then into
vocal or nonvocal sounds and finally into `excited' events (e.g. shouting,
running) or `normal' events (e.g. talking, walking).

SVMs are non-parametric classifiers
that can generalise well given a limited training set and have well defined
generalisation bounds~\cite{burges1998tutorial}. Port\^{e}lo et al.
\cite{portelo2009non} have experimented with an SVM when attempting to detect
events in film audio. They found it to perform well on some sounds
(aeroplanes, helicopters) but poorly on others (sirens, cats). By adding new
features and applying PCA they found better performance on siren sounds.
Moncrieff et al. \cite{moncrieff2001detecting} have also experimented with SVMs
when attempting to classify sound events in film audio. 
Chen et al. \cite{chen2006mixed} have used SVMs to classify audio data into
music, speech, environmental sounds, speech mixed with music and environmental
sounds mixed with music. Rouas et al.
\cite{rouas2006audio} have presented a comparison of SVMs and GMMs when
attempting to detect shouting in public transport vehicles. Testing on audio
recorded by actors on a train, they find that the SVM achieves a lower
false alarm rate whilst the GMM seems better at making positive identifications.

More recently, Deep Learning, i.e., artificial neural networks with more
than one intermediate hidden layer, have gained much popularity and have
achieved impressive results on several machine learning
tasks~\cite{lecun2015deep}. However, they do not seem to have been extensively
applied to AESR so far. Toyoda et al.~\cite{toyoda2004environmental} have used a
neural network to classify up to 45~different environmental sounds. They
achieved high recognition accuracies, though it should be mentioned that both
the training and test data sets used in their study may appear quite small in
comparison to the data sets used nowadays as benchmarks for speech recognition
and computer vision research. Baritelli et al.~\cite{beritelli2008pattern} used
a DNN to classify 10~different environmental sounds. The authors found that the
proposed system, which classified MFCCs directly, performed similarly to other
AESR systems which were requiring more complex feature extraction.
In~\cite{vidalsound}, the authors used neural networks to estimate the presence
or absence of people in a given recording, as well as the number of people and
their proximity to the recording device. Gencoglu et al.
\cite{gencoglu2014recognition}  applied a deep neural network to an
acoustic event detection task for classifying $61$ different classes of sounds.
Their results showed this classifier to outperform a
conventional GMM+HMM classification system.

Recently, the problem of AESR has been receiving more attention. Challenges like 
CLEAR \cite{temko2006clear} and DCASE \cite{stowell2015detection} have been proposed 
in order to benchmark the performance of different approaches and to stimulate 
research. But, if drawing a parallel between the evolution of AESR and
the evolution of speech recognition, the CLEAR and DCASE data sets
could be thought of as being at the stage of the TIMIT era \cite{zue1990speech}:
although the TIMIT data set was widely used for many years, it was limited in size
to about 5.4 hours of speech data. The subsequent growth in the volume of
available speech evaluation data sets has significantly contributed to
bringing speech recognition to the levels of performance known today,
where modern speech recognition systems are being trained over
1700~hours of data \cite{senior2015context}. 
AED research is following a similar trend where the domain is now facing a need to collect
larger data sets \cite{foster2015chime,Salamon:UrbanSound:ACMMM:14} in order to make larger scale evaluations
possible and to foster further technical progress.

Many studies explore 
feature learning to derive useful acoustic features that can be used to classify 
acoustic events \cite{salamon2015feature,salamon2015unsupervised}. A limited number of publications 
apply k-Nearest Neighbours (k-NNs)  \cite{sawhney1997situational,lukowicz2003soundbutton} 
or Random Forests \cite{stowell2014automatic} to AESR. However, these models have not been
included in our comparative study, which is restricted to GMM, SVM and DNN
models: these three types of models can be considered mainstream in terms of
having been applied more extensively to a larger range of audio classification
tasks which include speech, speaker and music classification and audio event
detection. In preliminary experiments, 
we observed that k-NN classifiers were clearly outperformed by the classifiers 
included in this study. 


\section{Computational cost \label{sec:cost}}

\subsection{Measure of computational cost used in this article}
The fundamental question studied in this paper is that of comparing various
machine learning algorithms {\em as a function of their computational cost}. A
corollary to this question is to compare the rate of performance degradation
that occurs when down-scaling various AESR algorithms by a comparable amount of
computing operations: a given class of algorithms might be more resilient to
down-scaling than another. While computational cost is related to computational
complexity, both notions are distinct: computational cost simply counts the
number of operations at a given model dimension, whereas computational
complexity expresses the mathematical law according to which the computational
cost scales up with the dimension of the input feature space. 

Computational cost is studied here at the sound recognition stage,
known as acoustic decoding. Unless specifically required by the
application, it is indeed fairly rare to see the training stage of machine
learning being realised directly on-board an embedded device. In most cases, it is
possible to design the training as an offline process realised on powerful
scientific computing platforms equipped with FPUs and GPUs. Offline training
delivers a floating point model which is then quantised according to the
needs of fixed point decoding in the embedded system \cite{smith1997scientist}.
It is generally accepted that the quantisation error introduced by this
operation is unlikely to affect the acoustic modelling performance \cite{gupta2015deep}.

The computational cost estimate used in this study simply counts four~types of
basic operations: addition, comparison, multiplication, and lookup table
retrieval (LUT).
{\em Multiply-add} operations are commonly found in dot products, matrix
multiplications and FFTs. Correlation operations, linear filtering or the
Mahalanobis distance, at the heart of many recognition algorithms, rely solely
on multiply-adds. Fixed point precision is fairly straightforward to manage
over this type of operation. {\em Divisions}, on the other hand, and in
particular matrix inversions, might be more difficult to manage. Regarding
matrix inversion, it can happen that quantisation errors add up and make the
matrix inversion algorithm unstable \cite{parhami2009computer}. In many cases, it is possible to
pre-compute the inversion of key parameters (e.g., variances), possibly offline
and in floating point before applying fixed-point quantisation. {\em Non-linear
operations} such as logarithms, exponentials, cosines, $N^{th}$~roots etc.
are required by some algorithms. Two approaches can commonly be taken to
transform non-linear functions into a series of multiply-adds:
either Taylor series expansion, or look-up tables (LUTs). LUTs are more memory
consuming than Taylor series expansions, but may require a lower number of
instructions to achieve the desired precision. Generally speaking, it is
important to be aware of the presence of a non-linearity in a particular
algorithm, before assessing its cost: algorithms which rely on a majority of
multiply-adds are more desirable than algorithms relying more heavily on
non-linearities.

For all the estimates presented below, it is assumed that these four~types
of operations have an equal cost. While this assumption would be true for a
majority of processors on the market, it may underestimate the cost of
non-linear functions if interpolated LUTs or Taylor series are used instead of
simpler direct LUT lookups. In addition to the 4 operations considered here, 
there are additional costs incurred with other operations (like data handling)
in the processor. However, we use the simplifying assumption that the
considered operations are running on the same core, thus minimizing data handling overhead.

As a case study, let us assume an attempt at running the K-nearest neighbours
algorithm on a Cortex-M4 processor running at a clock speed of 80MHz and
with 256kB of onboard memory. This is a very common hardware configuration for
consumer electronic devices such as video cameras. Assuming that 20\% of the processor
activity is needed for general system tasks means that
a maximum of 64 million of multiply or adds per second can be used for sound
recognition ($80\mbox{MHz} \times 80\% = 64\mbox{MHz}$).
Assuming sound sampled at 16kHz and audio feature extracted at, e.g., a window
shift of 256 samples, equivalent to 62.5 frames per second, means that the AESR
algorithm should use no more than 1,024,000 multiply or adds per analysis
frame (64MHz / 62.5Hz = 1\,024K instructions).

From there, assuming a K-Nearest neighbour algorithm with a 40 dimensional
observation vector $x$, mean vector $\mu$ and a Mahalanobis distance $( x - \mu ) \sigma^{-1} (x -
\mu)$ with a diagonal covariance matrix $\sigma$, there would be one~subtraction
and two~multiplications per dimension for the 
Mahalanobis distance calculation, plus one~operation for the accumulation across
each dimension: four~operations in total, thus entailing a maximum of 1,024,000
instructions / 4~operations / 40~dimensions = 6\,400 nearest neighbours on this
platform to achieve real-time recognition.

However, assuming that each nearest neighbour is a 40-dimensional vector of
32\,bits/2\,Bytes values, the memory requirement to store the model would be
512kB, which is double the 256kB available on the considered platform. Again
assuming that the code size occupies 20\% of the memory, leaving 80\% of the
memory for model storage, a maximum of $256\mbox{kB} \times 80\% \,/\,
40\,\mbox{features} \,/\, 2\mbox{Bytes} = 2\,560$ nearest neighbours only
which could be used as the acoustic model.

Depending on the choice of algorithm, memory is not always the limiting factor.
In the case of highly non-linear algorithms, the computational cost of achieving
real-time audio recognition might outweigh the model storage requirements.
While only the final integration can tell if the desired computational load and
precision are met, estimates can be obtained as illustrated in this
case study in order to forecast if an algorithm will fit on a particular
platform.

\begin{table*}[t!]
\newcommand\T{\rule{0pt}{2.6ex}}
\newcommand\B{\rule[-1.2ex]{0pt}{0pt}}
\renewcommand{\arraystretch}{1.3}
\begin{center}
\footnotesize
\begin{tabular}{| l | c | c | c | c |}

\hline
\textbf{Feature} & \textbf{Addition} & \textbf{Multiplication} & \textbf{
nonlinearity LUT lookup} \\
\hline
\hline

GMM & $2(M \cdot (D+1) + M)$ & $2M \cdot 2D$ & $M$ \\
SVM - Linear & $\lambda D + \lambda + 1$ & $\lambda \cdot D$ & 0 \\
SVM - Polynomial & $\lambda D +\lambda + 1$ & $\lambda(D+d)$ & 0 \\
SVM - Radial Basis Function & $2\lambda D + \lambda + 1$ & $\lambda(D+2)$ & $\lambda$ \\
SVM - Sigmoid & $\lambda D + \lambda +1$ & $\lambda(D+1)$ & $\lambda$ \\
DNN - Sigmoid & $H \cdot (1+D+L+(L-1)H)+1$ & $H \cdot (1+D+(L-1)H)$ & 
$L \cdot H + 1$\\
DNN - ReLU & $H \cdot (1+D+L+(L-1)H)+1$ & $H \cdot (1+D+(L-1)H)$ & 
$L \cdot H + 1$\\
RNN - Tanh & $H \cdot
(2+D+H+2(L-1)(H+1))+1$ & $H \cdot
(1+D+H+2(L-1)H)$ & $L \cdot H + 1$\\
\hline
\end{tabular}
\small
\caption{Computational cost per frame of each compared AESR algorithm. $D$ is
the dimensionality of the feature vector, $M$ is the number of Gaussian
mixtures for a GMM. $\lambda$ is the number of support vectors for a SVM, $d$
is the degree of a polynomial kernel. For the neural networks: $H$ is the
number of hidden units in each layer and $L$ is the number of layers.
\label{num_ops}
}
\end{center}
\end{table*}

\subsection{Computational cost estimates for the compared algorithms}
\label{sub:cost}
Most AED systems use a common 3-step pipeline \cite{stowell2015detection}. The audio
recording is first converted into  a time-frequency representation, usually
by applying the short-time Fourier transform (STFT)
to overlapping windows. This is followed by feature extraction from 
the time-frequency representation. Typically, Mel Frequency Cepstral
Coefficients (MFCCs) have been used as standard acoustic features since the seventies, due to favourable
properties when it comes to computing distances between sounds \cite{mermelstein1976}.
Then the acoustic features are input into an acoustic model to obtain a 
classification score, which in some cases is analogous to a posterior probability.
The classification score is then thresholded in order to obtain a classification decision.
Below, we provide an estimate of the computational cost for each of these operations.

%
\textbf{Feature extraction -}
Engineering the right feature space for AESR is essential,
since the definition of the feature space affects the separability of the
classes of acoustic data. Feature extraction from sound recordings thus forms
an additional step in the classification pipeline, which contributes to the
overall computational cost. In this study, the GMMs, SVMs and DNNs are trained 
on a set of input features that are 
typically used for audio and speech processing (Section \ref{sub:feature_extraction}). Although
recent studies demonstrate that neural networks can be trained  to jointly learn
the features and the classifier \cite{lecun2015deep}, we have found that this method is impractical for most
AESR problems where the amount of labelled data for training and testing is limited. Additionally, 
by training the different algorithms on the same set of features, we study the performance
of the various classifiers as a function of computation cost, without having to account
for the cost of extracting each type of feature from the raw audio waveform\footnote{For example in preliminary experiments, we observed that DNNs were able to achieve the same performance with log-spectrogram inputs as speech features (Section \ref{sub:feature_extraction}). However, in this case the GMMs and SVMs performed badly making it difficult to clearly compare performance.}. It also allows 
a fair comparison of the discriminative properties of different classification algorithms on 
the same input feature space. Therefore, we factor out the computational cost of 
feature extraction.

\textbf{Gaussian Mixture Models -}
Given a $D$ dimensional feature vector $x$, a GMM~\cite{reynolds2009gaussian} is
a weighted sum of Gaussian component densities which provides an estimate of the
likelihood of $x$ being generated by the probability distribution defined by
a set of parameters $\theta$:
\begin{equation}
p(x|\theta) = \sum\limits_{i=0}^{M-1} w_{i} \cdot g(x | \mu_{i}, \Sigma_{i} )
\label{eq:gauss1}
\end{equation}
where:
\begin{equation}
g(x | \mu_{i}, \Sigma_{i} ) = \frac{1}{(2\pi)^{\frac{D}{2}}
|\Sigma_{i}|^{\frac{1}{2}}} \cdot e^{ -\frac{1}{2} (x - \mu_i)'
\Sigma_{i}^{-1} (x - \mu_i)}
\label{eq:gauss2}
\end{equation}
where $\theta = (\mu_{i}, \Sigma_{i}, w_{i})_{0 \leq i < M}$ is the parameter
set, $M$ is the number of Gaussian components, $\mu_{i}$ are the means,
$\Sigma_{i}$ the covariance matrices and $w_{i}$ the weights of each Gaussian
component. In practice, computing the likelihood in the log domain avoids
numerical underflow. Furthermore, the covariance matrices are constrained to be
diagonal, thus transforming equations~(\ref{eq:gauss1}) and~(\ref{eq:gauss2}) into:
\begin{equation}
\begin{array}{l}
\log p(x|\theta) = \\
 \;\; \textrm{logsum}_{i=0}^{M-1} \left\{
\sum_{d=0}^{D-1}
\left(
x_d - \mu_{i,d}
\right)^2
\cdot
\sigma_{i,d}^{-1}
+ w_{i}
\right\}
+ K
\end{array}
\label{eq:gauss3}
\end{equation}
where $d$ is the dimension index, constant $K$ can be neglected for
classification purposes, and $\textrm{logsum}$ symbolises a recursive version of
function $\log(a+b) = \log a + \log(1 + e^{\left( \log b - \log a\right)})$ \cite{logsumexp},
which can be computed using a LUT on non-linearity $\log(1 + e^x)$.
Calculating the log likelihood of a $D$-dimensional vector given a GMM with
$M$~components therefore requires the following operations:
\begin{itemize}
\item $D$ subtractions (i.e., additions of negated means) and $2D$
multiplications per Gaussian to calculate the argument of the exponent.
\item $1$ extra addition per Gaussian to apply the weight in the log domain.
\item $M$ LUT lookups and $M$ additions for the non-linear $\textrm{logsum}$
cumulation across the Gaussian components.
\end{itemize}
This leads to the computational cost per audio frame expressed in the first
row of table~\ref{num_ops}.


\textbf{Support Vector Machines -}
\label{SVMs}
Support Vector Machines (SVMs) \cite{burges1998tutorial} are discriminative
classifiers. Given a set of data points belonging to two different
classes, the SVM determines the optimal separating hyperplane between the two
classes of data. In the linearly separable case, this is achieved by maximising
the margin between two hyperplanes that pass through a number of \emph{support
vectors}. The optimal separating hyperplane is defined by all points
$\textrm{x}$ that satisfy:
\begin{equation}
\textrm{x} \cdot \textrm{w} + b = 0 
\end{equation}
where $\textrm{w}$ is a normal vector to the hyperplane and $\frac{|b|}{\|w\|}$
is the perpendicular distance from the hyperplane to the origin. Given that all
data points $\textrm{x}_{i}$ satisfy:
\begin{equation}
\left\{
\begin{array}{ll}
\textrm{x}_{i} \cdot \textrm{w} + b \geq 1  & \mbox{for labels $y_{i} = +1$} \\
\textrm{x}_{i} \cdot \textrm{w} + b \leq -1 & \mbox{for labels $y_{i} = -1$}
\end{array}
\right.
\end{equation}
It can be shown that the maximum margin is defined by
minimising $\frac{\|\textrm{w}\|^2}{2}$. This can be solved using a Lagrangian
formulation of the problem, thus producing the multipliers $\alpha_{i}$ and the
decision function:
\begin{equation}
f(\textrm{x}) = \textrm{sgn} \Big( \sum\limits_{i=0}^{N-1} y_{i} \alpha_{i} \textrm{x} \cdot \textrm{x}_i + b \Big)
\end{equation}
where $N$ is the number of training examples and $\textrm{x}$ is a feature vector
we wish to classify. In practice, most of the $\alpha_{i}$ will be zero, and
the $\textrm{x}_i$ for non-zero $\alpha_{i}$ are called the \emph{support
vectors} of the algorithm.

In the case where the data is not linearly separable, a non-linear kernel function
$K(\textrm{x}_{i},\textrm{x}_{j})$ can be used to replace the dot products $\textrm{x} \cdot \textrm{x}_i$, 
with the effect of projecting the data into a higher dimensional space where it
could potentially become linearly separable. The decision function then becomes:
\begin{equation}
\label{SVM_classification}
f(\textrm{x}) = \textrm{sgn} \Big( \sum\limits_{i=0}^{N-1} y_{i} \alpha_{i} K(\textrm{x},\textrm{x}_{i}) + b \Big)
\end{equation}
Commonly used kernel functions include:

\medskip \noindent
\begin{tabular}{ll}
\hspace{-1.5ex} {$\vcenter{\hbox{\tiny$\bullet$}}$} Linear: &
\hspace{-2ex} $K(\textrm{x},\textrm{x}_{i}) = \textrm{x} \cdot \textrm{x}_{i}$
\\
\hspace{-1.5ex} {$\vcenter{\hbox{\tiny$\bullet$}}$} Polynomial: &
\hspace{-2ex} $K(\textrm{x},\textrm{x}_{i}) = (\textrm{x} \cdot
\textrm{x}_{i})^d$ \\
\hspace{-1.5ex} {$\vcenter{\hbox{\tiny$\bullet$}}$} Radial basis function (RBF):
& \hspace{-2ex} $K(\textrm{x},\textrm{x}_{i}) = \exp(-\gamma |
\textrm{x}-\textrm{x}_{i}|^2)$ \\
\hspace{-1.5ex} {$\vcenter{\hbox{\tiny$\bullet$}}$} Sigmoid: &
\hspace{-2ex} $K(\textrm{x},\textrm{x}_{i}) = \tanh(\textrm{x} \cdot
\textrm{x}_{i})$
\end{tabular}

\medskip \noindent
A further refinement to the SVM algorithm makes use of a \emph{soft margin}
whereby a hyperplane can still be found even if the data is non-separable
(perhaps due to mislabeled examples) \cite{burges1998tutorial}. The modified objective function is defined 
as follows:

\begin{equation}
\label{soft_margin}
\arg\min_{\textrm{w},\mathbf{\xi}, b } \left\{\frac{1}{2} \|\textrm{w}\|^2 + C
\sum_{i=1}^n \xi_i \right\}
\end{equation}

\vspace{-0.25em} \hspace{1cm} subject to: $ y_i(\mathbf{w}\cdot\mathbf{x_i} - b)
\ge 1 - \xi_i, \;\;\xi_i \ge 0$

\vspace{0.5em}
\noindent $\xi_i$ are non-negative \emph{slack variables}. The modified algorithm 
finds the best hyperplane that fits
the data while minimising the error of misclassified data points. The importance
of these error terms is determined by parameter~$C$,
which can control the tendency of the algorithm to over fit or under fit the data.

Given a support vector machine with $\lambda$ support vectors, a new
$D$-dimensional example can be classified using $\lambda$ additions to sum the
dot product results for each support vector, plus a further addition for the
bias term. For each dot product, a support vector requires the
computation of the kernel function, then $\lambda$ multiplications to apply
the $\alpha_i$ multipliers.

Some kernels require a number of elementary operations, whereas others require a
LUT lookup. Thus, depending on the kernel, the final costs are:


\begin{itemize}
\item SVM Linear -- $(\lambda \cdot D) + \lambda + 1$ additions and $(\lambda \cdot D) + \lambda$ multiplications.
\item Polynomial -- $\lambda(D + d + 2) + 1$ additions and $\lambda(D+2)$
multiplications, where $d$ is the degree of the polynomial.
\item Radial Basis Function -- $2\lambda D + \lambda + 1$ additions, $\lambda(D+2)$ multiplications and $\lambda$ exponential functions.
\item Sigmoid -- $\lambda(D+2) + 1$ additions and $\lambda(D+2)$ multiplications and $\lambda$ hyperbolic tangent functions.
\end{itemize}
The computational cost of SVMs per audio frame for these kernels is
summarized in Table~\ref{num_ops}.

\textbf{Deep Neural Networks -}

Two types of neural network architectures are compared in this study:
Feed-Forward Deep Neural Networks (DNNs) and Recurrent Neural Networks (RNNs).

A {\em Feed-Forward Deep Neural Network} processes the input by a series
of non-linear transformations given by:
\begin{equation}
\label{ffNN}
{h}^i = f \left( {W}^i {h}^{i-1}+{b}^i \right) \; \mbox{for} \; 0 \leq i \leq L
\end{equation}
where ${h}^0$ corresponds to the input $x$, ${W}^i,{b}^i$ 
are the weight matrix and bias vector for the $i^{th}$ layer, 
and the output of the final layer ${h}^L$ is the desired output. The
non-linearity $f$ in Equation~\ref{ffNN} is usually a sigmoid $f(x) =
1/\left( 1 + e^{-x} \right)$ or a hyperbolic tangent function $\tanh(x)$.
However recent research in various domains has showed that the Rectified Linear Unit
(ReLU)~\cite{glorot2011deep} non-linearity can lead to
faster convergence of models during training. The ReLU function is defined as
$f(x) = max(0,x)$, which is simpler than the sigmoid or $\tanh$ activation
because it only requires a piecewise linear operator instead of a fixed point sigmoid or
$\tanh$ LUT. 

While feed forward networks are designed for stationary data, {\em Recurrent
Neural Networks} (RNNs) are a class of neural networks designed for modelling
temporal data. RNNs predict the output $y_t$ given an input $x_t$, at some time $t$ in
the following way:
\begin{equation}
\label{RNN}
{h}^i_t = f \left( {W}^i_f {h}^{i-1}_t+{W}^i_r {h}^{i}_{t-1}+{b}^i \right) \;
\mbox{for} \; 0 \leq i \leq L
\end{equation}
where ${W}^i_f,{W}^i_r$ are the forward and recurrent weight matrices for the
$i^{th}$ recurrent layer, $b_i$ is bias for layer $i$,  and
$\theta = \left\{{W}^i_f,{W}^i_r,{b}^i \right\}_0^L$ are the model parameters.
The input layer $h^{0}_{t}$ corresponds to the input $x_t$ and the
output layer, $h^{L}_{t}$ corresponds to the output $y_t$. In Equation
\ref{RNN}, $f$ is the hyperbolic tangent function. The choice of $f$ for the
output layer depends on the particular modelling problem.

RNNs are characterised by the recurrent connections ${W}^i_r$ between the
hidden units. Similar to feed forward neural networks, RNNs can have several
recurrent layers stacked on top of each other to learn more complex
functions~\cite{graves2013speech}. Due to the recursive structure of the hidden
layer, at any time $t$, the RNN makes a prediction conditioned on the entire
sequence seen until time $t$: $y_t = f(x_0^t)$.

For simplicity, every neural network in this study was constrained to have the
same number $H$ of hidden units in each of its $L$ layers. The input dimensionality
is denoted by $D$. The forward pass through a feed-forward neural network therefore 
involves the following computations:

\begin{itemize}
\item Multiplying a column vector of size $n$ with a matrix 
of size $n \times k$ involves $n \cdot k$ additions and an equal number of 
multiplication operations. 
\item Therefore the first layer involves $\left( D \cdot H \right)$
multiplication operations and $H \cdot \left( D +1  \right)$ additions, where
the additional $H$ additions are due to the bias term. 
\item The remaining $L-1$ layers involve the following computations: $\left( L-1
\right) \cdot \left( H \cdot H\right)$ multiplications and $\left( L-1
\right) \cdot \left( H \cdot (H+1) \right)$ additions.
\item The output layer involves $H$ multiplications and $H+1$ additions. 
\item A non-linearity is applied to the outputs of each layer, leading to a 
total of $L \cdot H+1$ non-linearities.  
\item The RNN forward pass includes an additional matrix multiplication due to the
recurrent connections. This involves $H \cdot H$ multiplications and an equal 
number of addition operations. 
\end{itemize}
The computational cost estimates for DNNs and RNNs are summarised in Table~\ref{num_ops}.

\section{Performance metrics for sound recognition \label{sec:perf}}
Classification systems produce two types of errors: False Negatives (FN), a.k.a. Missed Detections (MD),
i.e., target sounds which have not been detected, and False
Positives (FP), a.k.a. False Alarms (FA), i.e., non-target sounds which have been
mistakenly classified as the target sound~\cite{murphy2012machine}. 
Their correlates are the True Positives (TP), or
correctly detected target sounds, and True Negatives (TN), or correctly
rejected non-target sounds. In our experiments, ``sounds'' refer to feature
frames associated with a particular sound class label.
The above metrics are usually expressed as percentages of the available
testing samples. They can be computed \emph{after an operating point
or threshold} has been selected \cite{japkowicz2011evaluating} for a given classifier. 
Indeed, most classifiers output a real valued \emph{score}, which can be, e.g., a
log-likelihood ratio, a posterior probability or a distance. The binary
target/non-target classification is then made by comparing the score to a
threshold. Thus, a system can be made either more permissive, by choosing a
threshold which lets through more True Positives at the expense of
generating more False Positives, or the system can be made more conservative,
if the threshold is set to filter out more False Positives at the expense of
rejecting more sounds in general and thus generating more Missed Detections.
For example, the operation point of a GMM classifier is defined by choosing a
log-likelihood threshold.

Ideally, AESR systems should be compared independently of the choice of threshold.
This can be achieved through the use of
Detection Error Tradeoff (DET) curves~\cite{martin1997det}, which depict the
FA/MD compromises achieved across a sweep of operation points. DET curves are
essentially equivalent to Receiver Operating Curves
(ROC)~\cite{martin1997det}, but plotted into normal deviate scale to
linearise the curve, under the assumption that FA and MD scores are normally
distributed. Given that the goal is to minimise the FA/MD compromises globally
across all possible operation points, a DET curve closer to the origin means a
better classifier. DET curve performance can thus be summarised for convenience
into a single figure called the Equal Error Rate (EER), which is the point
where the DET curve crosses the \%FA = \%MD diagonal line. In other terms, the
EER represents the operation point where the system generates an equal rate of
FA and MD errors. 

For GMM based classification systems, the output score is the ratio between the likelihood
of the target GMM and the likelihood of the universal background model (UBM): $P(x|\theta_{\mbox{\tiny
target sound}})/P(x|\theta_{\mbox{\tiny UBM}})$ \cite{reynolds2009gaussian}.
For neural networks, the output layer consists in a single unit followed by a
sigmoid function, thus yielding a real-valued class membership
probability $P(C=1|x)$ taken as the score value.  For SVMs, the score can be
defined as the argument of the $\textrm{sgn}()$ function in
Equation~\ref{SVM_classification}: whereas the $\textrm{sgn}()$ function always
amounts to classifying against a threshold fixed to zero, using its argument as
a score suggests a more gradual notion of deviation from the margin defined by
the support vectors, or in other terms it suggests a more gradual measure of
belonging to one side or the other of the margin.

\section{Data sets and methodology \label{sec:methodology}}

\subsection{Data sets}

For the problem of AED, the training data set consists of audio recordings along with corresponding
annotations or ground truth labels. The ground truth labels are usually provided in 
the form of alignments, i.e. onset times, offset times and the audio event
labels. This is in contrast to ASC problems where the ground truth does not include alignments. 
ASC is similar to the problem of speech recognition where each recording has an 
associated word-level transcription. The alignments are then obtained as a by-product of the
speech recognition model. For the problem of AED, annotating large quantities of data with 
onset and offset times requires significant effort from human annotators.

Historically, relatively small data sets of labelled audio have been used for evaluating AED systems.
As a matter of fact, the two most popular benchmarks for AED algorithms known to date
are the DCASE challenge \cite{stowell2015detection} and the CLEAR challenge \cite{temko2006clear}. 
The DCASE challenge data provides 20 examples for each of the 16 event classes. While 
the data set for the CLEAR challenge contains approximately 60 training instances for 
each of the 13 classes. The typical approach to AED so far has been to extract features from the audio
and then use a classifier or an \emph{acoustic model} to classify the frames into event labels.  
Additionally, the temporal relationship between acoustic frames and outputs is modelled 
 using HMMs \cite{stowell2015detection}. 
 
In this study, we investigate whether a data-driven approach to AED can yield improved performance. 
Inspired by the recent success of Deep Learning \cite{lecun2015deep}, we investigate whether neural network
acoustic models can outperform existing models when trained on sufficiently large data sets. 
Deep Neural Networks (DNNs) have recently contributed to significant progress in the fields 
of computer vision, speech recognition, natural
language processing and other domains of machine learning. The superior performance of neural
networks can be largely attributed to more computational power and the availability of large quantities
of labelled data. Neural network models with millions of parameters can now be trained on 
distributed platforms to achieve good generalisation performance \cite{szegedy2015going}. However, 
despite the strong motivation to use neural networks for various tasks, their application
is limited by the availability of labelled data. For example, we trained many DNN architectures
on the DCASE data, but were unable to beat the performance of a baseline GMM system.

In order to train large neural network models, we perform experiments using 
three private data sets made available 
by Audio Analytic Ltd.\footnote{ {\tt \url{http://www.AudioAnalytic.com}} \\
The data can be made available to selected research partners 
upon setting suitable contractual agreements.} 
to support this study. These three data sets are subsets of much larger data 
collection campaigns led by Audio Analytic Ltd. to support the
development of smoke alarm and baby cry detection products. Ground truth
annotations with onset and offset times obtained from human annotators were
also provided. The motivation for using
these data sets is twofold. Firstly, to provide a large number of training and test 
examples for training models that recognize baby cries and smoke alarm sounds. 
Secondly, a very large \emph{world} data set of ambient sounds is used as a source of 
impostor data for training and testing, in a way which emulates the quasi infinite
prior probability of the system being exposed to non-target sounds instead of target ones.
In other words, there are potentially thousands of impostor sounds which
the system must reject in addition to identifying the target sounds correctly. While 
the DCASE challenge evaluates how an AED system performs at detecting one sound in the presence
of 15 impostors or non-target sounds, the world data set allows to evaluate the
performance of detecting smoke alarms or baby cries against the presence of over 1000 impostor
sounds, in order to reflect a more realistic use case.



The \emph{Baby Cry} data subset comprises a
proportion of baby cries obtained from two different recording conditions:
one through camcorder in a hospital environment, one through uncontrolled 
hand-held recorders in uncontrolled conditions (both indoors and outdoors). 
Each recording was from a different baby. The train/test split was done evenly, 
and set to achieve the same balance of both conditions in both the train and 
test set (about 3/4 hospital condition and 1/4 uncontrolled condition) without
any overlap between sources for training and testing. The recordings
in this data set sum up to $4\,822$ seconds of training data and $3\,669$
seconds of test data, from which $224\,076$ feature frames in the training set
and $90\,958$ feature frames in the test set correspond to target baby cry
sounds.

The \emph{Smoke Alarm} data subset comprises recordings of 10 smoke
alarm models, recorded through 13 different channels across 3 British homes. 
While it may be thought that smoke alarms are simple sounds, the variability 
associated with different tones, different audio patterns, different room 
responses and different channel responses does require some generalisation 
capability which either may not be available in simpler fingerprinting or 
template matching methods, or would require a very large database of templates.
The training set comprises recordings
from two of the homes, while the test set is composed of recordings from the 
remaining home. Thus, while the same device can appear in
both the training and evaluation sets, the recording conditions are always different.
So the results presented over the smoke alarm set focus mainly on the generalisation
across room responses and recording conditions.
The data set provided for
this study consists of $15\,271$ seconds of training data and $5\,043$~seconds
of testing data, of which $194\,142$ feature frames in the training set and
$114\,753$ feature frames in the test set correspond to target smoke alarm
sounds.

The \emph{World} data set contains about 900 files covering 10 second samples
of a wide variety of complex acoustic scenes recorded
from numerous indoor and outdoor locations around the UK, across a wide and
uncontrolled range of devices, and potentially covering more than a thousand
sound classes. For example, a 10 seconds sample of “train station” scene from
the world data set covers, e.g., train noise, speech, departure whistle and more,
while a 10 second sample of “supermarket” covers babble noise, till beeps and
children’s voices. While it does not seem feasible to enumerate all the specific
classes from the World set, assuming that there are more than two classes
per file across the 410 files reserved for testing leads to
an estimate of more than a thousand classes in the test set.
The 500 other files of the world set are used to train a non-target model,
i.e., a Universal Background Model (UBM)~\cite{reynolds2009gaussian}
in the case of GMMs, or a negative input for neural network and SVM training.
It was ensured that the audio files in the both the training and testing world
sets were coming from completely disjoint recording sessions. The World data set provided
for this study consists of $5\,000$~seconds of training data and $4\,089$ seconds of test
data, thus amounting to $312\,500$ feature frames for training and $255\,562$
feature frames for testing. 

Where necessary, the recordings were converted to 16kHz, 16~bits using a high
quality resampling algorithm. No other pre-processing was applied to the
waveform before feature extraction.

\subsection{Feature Extraction \label{sub:feature_extraction}}
For this experiment, Mel-frequency cepstral coefficient (MFCC) features were
extracted from the audio. MFCC features are extensively used in speech
recognition and in environmental sound recognition \cite{clavel2005events,valenzise2007scream,
portelo2009non,radhakrishnan2005audio}. In addition to MFCCs, the
spectral centroid \cite{clavel2005events}, spectral flatness
\cite{portelo2009non}, spectral rolloff \cite{valenzise2007scream}, spectral
kurtosis \cite{valenzise2007scream} and zero crossing rate
\cite{valenzise2007scream,atrey2006audio} were also computed. With audio
sampled at 16kHz, all the features were calculated with a window size of
$512$~samples and a hop size of $256$~samples. The first $13$~MFCCs
were used for our experiments. Concatenating all the features thus lead
to an $18$ dimensional feature vector of real values. Temporal
information was included in the input representation by appending the first and
second order differences of each feature. This led to an input representation with
$18 \times 3 = 54$~dimensions. It should be noted that the temporal difference features were
used as inputs to all models except the RNNs, where only the $18$~dimensional
features were used as inputs. This is due to the fact that the RNN is designed to
make predictions conditioned on the entire sequence history (Equation \ref{RNN}). 
Therefore, explicitly adding temporal information to the inputs is unnecessary.
The reduction in input dimensionality and related reduction in computational
cost is offset by the extra weight matrix in the RNN (Table~\ref{num_ops}). 
For all the experiments, the data was normalised to have zero mean and
unit standard deviation, with the mean and standard deviation calculated 
over the entire training data independently for each feature dimension.

\subsection{Training Methodology \label{training}}

\textbf{Gaussian Mixture Models -} GMMs were trained using the expectation
maximisation algorithm (EM) \cite{bishop2006pattern}. The covariance matrices were
constrained to be diagonal\footnote{In preliminary experiments, we observed that
a full covariance matrix did not yield any improvement in performance.}. 
A grid search was performed to find the optimal
number of Gaussians in the mixture, on the training set. The best model with number 
of components $M \in \{ 1,2,4,8,16,32,64,128,256,512,1024 \}$ was used for evaluation. A validation set was 
created by randomly selecting 20\% of the training data, while the remaining 80\% of the data
was used for parameter estimation. The model that performed the best on the validation set
was later used for testing. All the Gaussians were initialised 
using the k-means++ algorithm \cite{arthur2007k}. 
A single UBM was trained on the World training data set. One GMM per target class was then trained
independently of the UBM. We also performed experiments where GMMs for each 
class were estimated by adapting the UBM by Maximum A Posteriori (MAP) adaptation \cite{reynolds2000speaker}.
We observed that this method yielded inferior results. We also observed worse results when no UBM
was used and class membership was determined by thresholding the outputs of individual 
class GMMs. In the following, results are therefore only reported for the
best performing method, which is the likelihood ratio between independently
trained UBM and target GMMs.

\textbf{Support Vector Machines -} The linear, polynomial, radial basis function
(RBF) and sigmoid kernels were compared in our experiments. In addition to
optimising the kernel parameters ($d,\gamma$), a grid search was performed over the penalty
parameter $C$ (Equation \ref{soft_margin}). In practice, real-world data are often 
non-separable, thus a soft margin has been used as necessary.
Model performance was estimated by evaluating
performance on a validation set. The validation set was formed by randomly selecting 20\%
of the training data, while the remaining data was used for parameter estimation. Again, the 
best performing model on the validation set was used for evaluation. 
SVM training is known not to scale well to very large
data sets~\cite{burges1998tutorial}: it has a time-complexity of $O(T^3)$, where
$T$ is the number of training examples. As a matter of fact, our preliminary
attempts at using the full training data sets led to prohibitive training times.
Standard practice in that case is to down-sample the training set by randomly
drawing $T$ frames from the target set, and the same number $T$ of frames from
the world set as negative examples for training. Results presented in
section~\ref{sec:results} with $T \in \left\{ 500,2000 \right\}$ show that
contrary to the intuitive thought that down-sampling might have disadvantaged
the SVMs by reducing the amount of training data compared to the other
machines, such data reduction actually improves SVM performance. As such,
down-sampling can be thought of as an integral part of the training process,
rather than as a reduction of the training data.

\textbf{Deep Neural Networks -} All neural network architectures
were trained to output a distribution over the presence or absence of a class,
using the backpropagation algorithm and stochastic gradient descent (SGD)
\cite{lecun2012efficient}. The training data was divided into a $80/20$
training/validation split to find the optimum training hyper-parameters. All the
target frames were used for each class, and an equal number of non-target frames
was sampled randomly from the World data set, for both training and validation.

For the DNNs, a grid search was performed over the following parameters: number
of hidden layers $L \in \left\{ 1,2,3,4 \right\}$, number of hidden units per layer $H \in
\left\{ 10,25,50,100,150 \right\}$, hidden activations $act \in \left\{
sigmoid, ReLU \right\}$. In order to minimise parameter tuning, we used
ADADELTA \cite{zeiler2012adadelta} to adapt the learning rate over iterations.
The networks were trained using mini-batches of size $100$ and training was
accelerated using an NVIDIA Tesla K40c GPU. A constant dropout rate
\cite{srivastava2014dropout} of $0.2$ was used for all layers. The training
was stopped if the cost on the validation set did not decrease after $20$
epochs.

For the RNNs, a grid search was performed over the following parameters: 
number of hidden layers $L \in \left\{ 1,2,3 \right\}$ and number of hidden units 
per layer $H \in \left\{10,25,50,100,150 \right\}$. An initial learning rate of $0.001$ was used and
linearly decreased to $0$ over $1000$ iterations. A constant momentum
rate of $0.9$ was used for all the updates. The training was stopped if the
error on the validation set did not decrease after $20$ epochs. The training
data was further divided into sub-sequences of length $100$ and the networks
were trained on these sub-sequences without any mini-batching. Gradient clipping
\cite{bengio2013advances} was also used to avoid the exploding gradient problem
in the early stages of RNN training. Clipping was triggered if the norm of the
gradient update exceeded $10$.

\section{Experimental results \label{sec:results}}

\begin{figure*}[ht]

\begin{center}
\begin{tabular}{ccc}

\subfloat[Baby Cry data set]{\includegraphics[width=0.42\textwidth,height=0.42\textwidth]{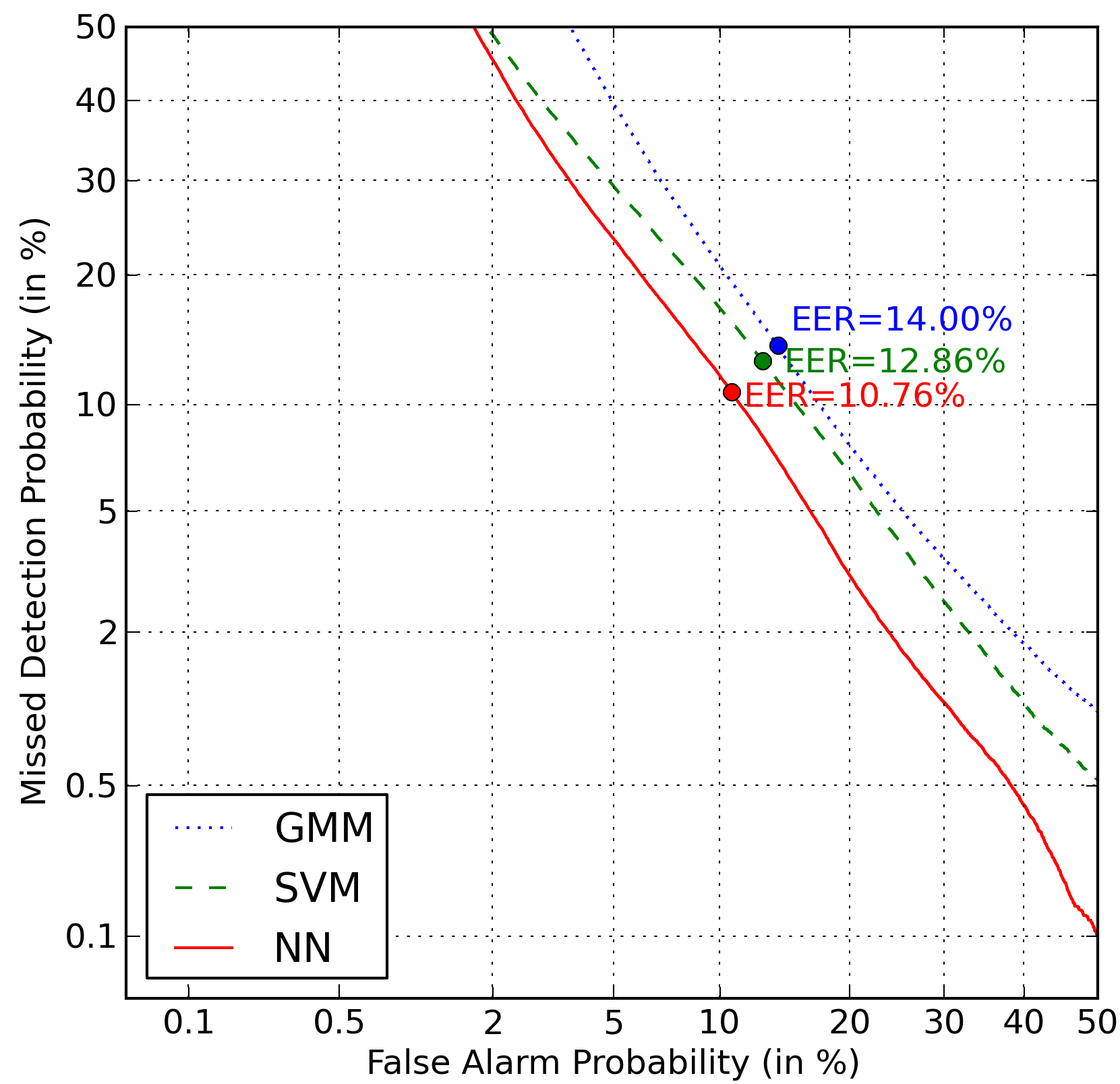}\label{DET_baby_cry}}

&

\hspace{1cm}
 
&

\subfloat[Smoke Alarm data set]{\includegraphics[width=0.42\textwidth,height=0.42\textwidth]{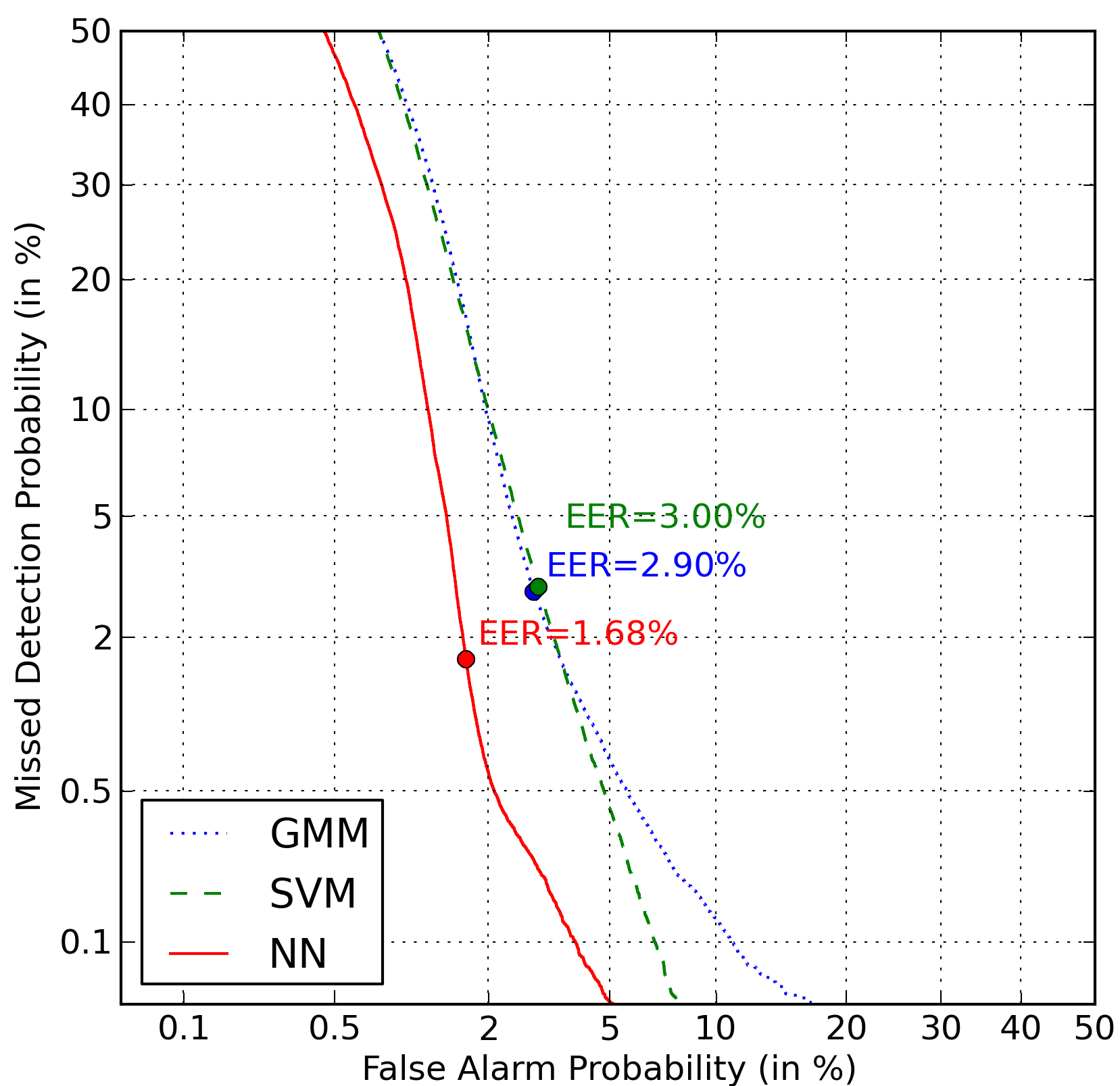}\label{DET_smoke_alarms}}\\
\end{tabular}
\end{center}
\caption{DET curves comparing frame classification performance of the acoustic classifiers.}
\label{DET_curves}
\end{figure*}

\begin{table*}
\begin{center}

\begin{tabular}{ccc}

\scalebox{1.}{
\subfloat[]{
\begin{tabular}{| l | c | c |}
\hline
\textbf{Best Baby Cry classifiers} & \textbf{EER} & \textbf{\# Ops.} \\
\hline
GMM, $M=32$  & 14.0 & 10\,560 \\
\parbox[t]{28ex}{Linear SVM, $T=2000$, \\ $C=1.0$, $\lambda=655$} & 12.9 &
101\,985
\\
\parbox[t]{28ex}{Feed-forward DNN, sigmoid, \\ $L=2$, $H=50$} & 10.8 &
10\,702 \\
\hline
\end{tabular}
\label{performance_baby_cry}
}
}
&

\hspace{1cm}
 
&
\scalebox{1.}{
\subfloat[]{
\begin{tabular}{| l | c | c |}
\hline
\textbf{Best Smoke Alarm classifiers} & \textbf{EER} & \textbf{\# Ops.} \\
\hline
GMM, $M=16$  & 2.9 & 5\,280 \\
\parbox[t]{28ex}{Linear SVM, $T=2000$, \\ $C=0.1$, $\lambda=152$} & 3.0 &
46\,655
\\
\parbox[t]{28ex}{Feed-forward DNN, sigmoid, \\ $L=2$, $H=25$} & 1.7 & 4\,102 \\
\hline
\end{tabular}
\label{performance_smoke_alarms}
}
}

\end{tabular}
\caption{Performance of the best classifiers on the Baby Cry and Smoke Alarm data sets.}
\label{classifier_performance}
\end{center}
\end{table*}

\begin{figure*}[t!]
\subfloat[Baby Cry Data set]{\includegraphics[width=0.5\textwidth,height=0.45\textwidth]{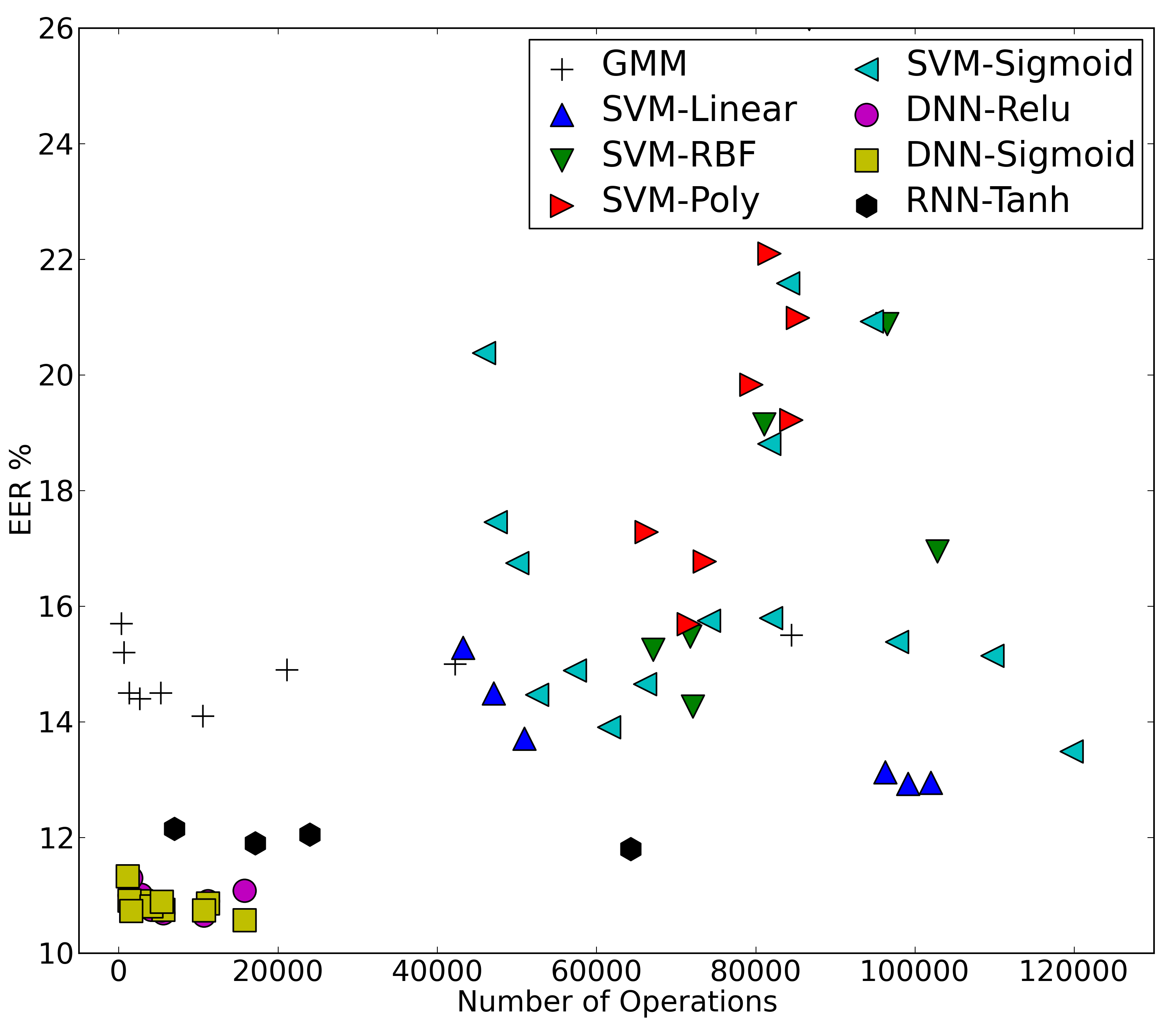}\label{baby_cry_scatter}}
\hfill
\subfloat[Smoke Alarm Data set]{\includegraphics[width=0.5\textwidth,height=0.45\textwidth]{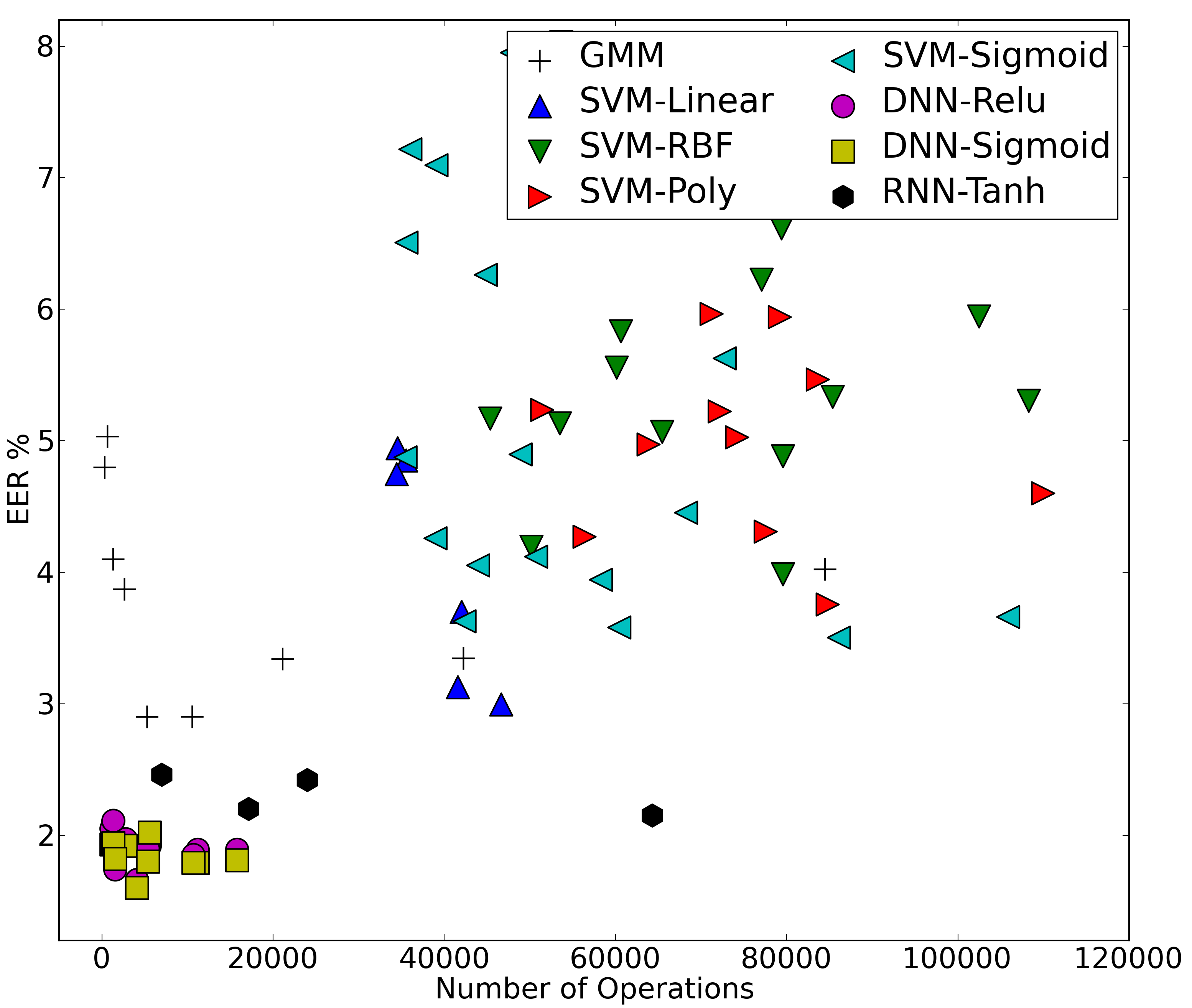}\label{smoke_alarm_scatter}}
\caption{Acoustic frame classification performance (EER percentage) as a
function of the number of operations per frame, for each of the tested models,
across the Baby Cry and Smoke Alarms data sets. }
\label{scatter_plots}
\end{figure*}

In this section, classification performance is analysed for the task of recognising
smoke alarms and baby cries against the large number of impostor sounds from the world set.
Table \ref{classifier_performance} presents the EER for the best performing classifiers of each
type and their associated computational cost. Figure \ref{DET_curves} shows the corresponding DET 
curves. In Figure \ref{scatter_plots} we present a more detailed
comparison between the performance of various classifier types against the computational 
cost involved in classifying a frame of input at test time. 

\textbf{Baby Cry data set -} From Table \ref{performance_baby_cry} we observe that the 
best performing GMM has 32 components and achieves an EER of 14.0\% for frame-wise 
classification. From Figure \ref{baby_cry_scatter} (+ markers) we observe that there is no performance 
improvement when the number of Gaussian components is increased beyond 32.
It should be noted that the best performing GMM (with $32$ components) has lower computational cost compared to 
any SVM classifier, and a cost comparable on average with DNN classifiers.

From Table \ref{performance_baby_cry}, we observe that a SVM with a linear kernel 
is the best performing SVM classifier, with an EER of $12.9\%$. The SVM was trained on
2000 examples, resulting in 655 support vectors. From Figure \ref{baby_cry_scatter}, we 
note that the performance of the linear SVM (blue triangles) increases as the computational complexity 
(number of support vectors) is increased. The number of support vectors can also
be controlled by varying the parameter $C$. For the best performing linear SVM, $C=1.0$ yields the best 
results. We observe that SVMs with a sigmoid kernel (light blue triangles) yield similar results to the linear SVM 
(Figure \ref{baby_cry_scatter}). The best SVM with a sigmoid kernel yields an EER of 13.5\%, while 
the second best sigmoid SVM has an EER of 13.9\%. As in the linear case, we observe 
an improvement in test performance as the computational cost or the number 
of support vectors is increased.  

From Figure \ref{baby_cry_scatter}, we note that SVMs with RBF 
and polynomial kernels (green and red triangles) are outperformed by linear and sigmoid SVMs, both
in terms of \%EER and computational cost. There is also no observable trend between
test performance and computational cost. For the polynomial SVM (red triangles), we found a 
kernel with $d = 3$ yielded the best performance, while low values of gamma 
$\gamma \in (0.005,0.01)$ provided the best results for the RBF kernel (Section \ref{SVMs}).  

The number of training examples $T = 2000$ was empirically determined to be the
optimal value: adding more training examples did not yield any improvement in
performance. Conversely, we tried training the same classifiers with $T
= 500$ training examples, since the computational cost of SVMs is determined
both by the type of kernel and the number of support vectors, the latter being
controlled by varying the number of training examples and the penalty parameter
$C$. With $T = 500$ and $C=0.1$, we were able to
achieve a minimum EER rate of $13.7\%$ with the linear SVM, while halving the number of operations.

From Table \ref{performance_baby_cry}, the best performing neural network
architecture, which achieves an EER of $10.8\%$, is a feed-forward DNN with
sigmoid activations for the hidden units. Figure \ref{DET_baby_cry} shows that
the neural network clearly outperforms all the other models. From Figure \ref{baby_cry_scatter} we observe that the 
feed forward DNNs (circle and square markers) achieve similar test performance over a wide range of computational
costs. This demonstrates that the network performance is not particularly sensitive to the 
specific number of hidden units in each layer. However, we did observe that
networks which were deeper ($>1$ hidden layer) yielded better performance. From Figure \ref{baby_cry_scatter}, 
we observe that the RNN architectures (hexagonal markers) yield slightly worse performance and are computationally
more expensive. An interesting observation from Figure \ref{baby_cry_scatter} is that 
feed-forward DNNs with both sigmoid and ReLU activations yield similar results. This is 
a very important factor when deploying these models on embedded hardware, since a ReLU net
can be implemented with only linear operations (multiplications and additions), without 
the need for costly Taylor series expansions or LUTs.

\textbf{Smoke Alarm data set -} From Table \ref{performance_smoke_alarms}, we
observe that the best performing GMM yields an EER of 2.9\% and uses 16 mixture
components. From Figure~\ref{smoke_alarm_scatter} (+ markers), we observe that the GMM
performance improves as the computation cost (number of Gaussians) increases till $M=32$.
Beyond this, increasing the number of Gaussian components does not improve results. 
Again, we observe that the best performing GMM with $M=32$ has much lower computational
cost compared to SVMs, and a cost comparable on average with DNNs.

Similar to the results on the Baby Cry data set, the linear and sigmoid
kernel SVMs show the best performance, out of all four SVM kernel types. 
The best linear SVM has an EER of 3.0\% -- a small
improvement over the best GMM approach, but with a small increase in the number
of operations. The model used 2000 training examples and $C=1.0$. From Figure 
\ref{smoke_alarm_scatter} we again observe an improvement in performance, with 
an increase in computation cost for both the linear and sigmoid kernels (blue triangles). 
The best sigmoid kernel SVM scored 3.5\% with 2000 training examples, while
another configuration scored 3.6\% with $T = 500$ and $C = 1$, at half the
number of operations. Again, the polynomial and RBF kernels (red and green triangles) yield lower performance, 
with no observable trend in terms of performance versus computational cost 
(Figure \ref{smoke_alarm_scatter}). 

From Table \ref{performance_smoke_alarms}, we observe that a feed-forward sigmoid DNN
 yields the best performance, with an EER of $1.6\%$. 
From the DET curves (Figure \ref{DET_smoke_alarms}), we see that the neural network
clearly outperforms the other models. From figure \ref{smoke_alarm_scatter} we note
that the neural networks consistently perform better than the other models, over a 
wide range of computational costs, which correspond to different network configurations 
(number of layers, number of units in each layer). The ReLU networks perform similarly 
to the sigmoid networks, while the RNNs perform worse and are computationally more
costly. 

It is worth noting that the performance of all classifiers is significantly better for the smoke alarm 
sounds, since the smoke alarms are composed of simple tones. On the other 
hand, baby cries have a large variability and are therefore more difficult to classify.

\section{Conclusion \label{sec:conclusion}}
%

In this study, we compare the performance of neural network acoustic models with
GMMs and SVMs on an environmental audio event detection task. Unlike other machine learning systems,
AESR systems are usually deployed on embedded hardware, which imposes many computational constraints.
Keeping this in mind, we compare the performance of the models as a function of their computational 
cost. We evaluate the models on two tasks, detecting baby cries and detecting smoke alarms against
a large number of impostor sounds.
These data sets are much larger than the popular data sets found in AESR literature, which
enables us to train neural network acoustic models. Additionally, the large
number of impostor sounds allows to investigate the performance of the proposed 
models in a testing scenario that is closer to practical use cases than previously
available data sets.

Results suggest that GMMs provide a low cost baseline for classification, across
both data sets. The GMM acoustic models are able to perform reasonably well at 
a modest computational cost. SVMs with linear and sigmoid kernels yield similar 
EER performance compared to GMMs, but their computational cost is overall higher.
The computational cost of the SVM is determined by the number of support vectors. 
Unlike GMMs, SVMs are non-parametric models which do not allow the direct specification
of model parameters, although the number of support vectors can be indirectly controlled
with regularisation. Finally, our results suggest that deep neural
networks consistently outperform both the GMMs and the SVMs on both
data sets. The computational cost of DNNs can be controlled by
limiting the number of hidden units and the number of layers. While changes in
the number of units in the hidden layers did not appear to have a large impact
on performance, deeper networks appeared to perform better in all cases.
Additionally, neural networks with ReLU activations achieved good performance,
while being an attractive choice for deployment on embedded devices because
they do not require expensive LUT lookup operations.

In the future, we would like to expand the evaluations presented here to include more
event classes. However, the lack of large data sets for AESR problems is a major limitation. 
We hope that studies like this one will encourage collaboration with industry partners to collect 
large data sets for more rigorous evaluations. We would also like to investigate the performance
of acoustic models as a function of the memory efficiency, since memory is an important 
consideration when designing models for embedded hardware. 

\section{Acknowledgements}

The authors would like to thank the three anonymous reviewers whose comments helped to improve the article.

This work was supported by funding from Innovate UK, EPSRC grants EP/M507088/1 \& EP/N014111/1 from the UK Engineering and Physical Sciences Research Council, as well as private funding from Audio Analytic Ltd.

\ifCLASSOPTIONcaptionsoff
  \newpage
\fi



%

\bibliographystyle{IEEEtran}
\bibliography{main}

\end{document}